\documentclass[12pt]{amsart}
\usepackage{amscd,amsmath,amsthm,amssymb,graphics}
\usepackage{amsfonts,amssymb,amscd,amsmath,enumitem,verbatim}
\usepackage{hyperref}
\usepackage{authblk}
\usepackage{blindtext}
\theoremstyle{plain}
\newtheorem{Theorem}{Theorem}
\newtheorem{Lemma}[Theorem]{Lemma}

\theoremstyle{definition}
\newtheorem{Definition}[Theorem]{Definition}
\newtheorem{Remark}[Theorem]{Remark}

\usepackage{float}
\begin{document}
\title[Special subsets of addresses for blockchains]{Special subsets of addresses for blockchains using the secp256k1 curve}

\maketitle
\begin{center}
  ANTONIO J. DI SCALA\textsuperscript{1}, ANDREA GANGEMI\textsuperscript{1},
  GIULIANO ROMEO\textsuperscript{1} AND
  GABRIELE VERNETTI\textsuperscript{2} \par \bigskip

  \textsuperscript{1} DISMA, Department of Mathematical Sciences, Politecnico of Turin \par
  \textsuperscript{2} DAUIN, Department of Control and Computer Engineering, Politecnico of Turin \par
   \bigskip
   \end{center}

\begin{abstract}
In 2020 Sala, Sogiorno and Taufer \cite{SALA} have been able to find the private keys of some Bitcoin addresses, thus being able to spend the cryptocurrency linked to them. This result was unexpected, since the recovery of non-trivial private keys for blockchain addresses is deemed to be an infeasible problem. In this paper we widen this analysis by mounting a similar attack to other small subsets of the set of private keys. We then apply it  to other blockchains as well, examining Ethereum, Dogecoin, Litecoin, Dash, Zcash and Bitcoin Cash. In addition to the results, we also explain the techniques we have used to perform this exhaustive search for all the addresses that have ever appeared in these blockchains.
\end{abstract}

\section{Introduction}
The impact that Bitcoin has had on modern society hardly needs to be explained. In just a few years since the publication of Satoshi Nakamoto's white paper \cite{WP}, blockchain technology has taken hold all over the world. Its main objective is the creation of an open, public, decentralised and immutable ledger whose reliability is not based on a trusted third-party. It is natural that a tool with these features could find a use in several other fields besides transactions validation. Some references on the various fields where the blockchain technology has been employed can be found in \cite{BA}.\\
The blockchain that has catched most of the interest after Bitcoin is perhaps Ethereum, theorised in 2013 by Vitalik Buterin \cite{WPE}. The aim of Ethereum is to be a versatile world programmable computer in which smart contracts, i.e. decentralised programs written in a Turing-complete programming language, can be executed. A mathematical description of Ethereum's blockchain has been published by Gavin Wood in the yellow paper \cite{YP}.\\
In a blockchain, the security of transactions is guaranteeed by public-key cryptography and, in particular, by the difficulty of solving the discrete logarithm problem. Using the best known algorithms it is practically impossible, in a general case, to recover the private key starting from the knowledge of the public address. For all these reasons, it has been quite unexpected when, in 2020, Sala, Sogiorno and Taufer \cite{SALA} found the private keys of some existing Bitcoin addresses, being able to spend cryptocurrencies on their behalf. They mounted an attack on a small multiplicative subgroup of a group that is mapped to the group of points of Bitcoin's elliptic curve \textit{secp256k1}. For this small set, a brute-force attack has been feasible and, against any odds, $4$ of those addresses coincided with some actually used in Bitcoin's history.\\
The same authors left open the problem of understanding the reasons for such pathological behavior and the analysis of other cryptocurriences together with other small algebraic structures.\\

In this paper we widen the range of the inspection to other $7$ subsets of the same order, obtained as cosets of the group used in \cite{SALA}. Furthermore, we perform the same examination  for Ethereum addresses and for  some other famous cryptocurrencies that share the same elliptic curve: Dogecoin \cite{DOGE}, Litecoin \cite{LITE}, Zcash \cite{ZEC}, Dash \cite{DASH} and Bitcoin Cash \cite{BCH}.
The paper is organised as follows: in Section 2 we list some basic definitions useful for the following, in Section 3 we define the small subsets we have decided to analyse, in Section 4 we summarise the address generation algorithms used by the blockchains we have chosen and in Section 5 we explain the strategies used to extract the complete list of addresses. Finally, in Section 6 we report the results we have obtained and we make a comparison with other elliptic curves that we find particularly interesting.

\vskip1cm
\section{Preliminaries}\label{sect2}
Let us recall some basic definitions from algebra and, in particular, from group theory and  elliptic curves over finite fields.

\subsection{Group theory}

\begin{Lemma}\label{cyclic}
Let us consider a cyclic group $G=\langle g\rangle $ of order $n\in\mathbb{N}$. Then for each divisor $d$ of $n$ there exists a unique subgroup $H$ of order $d$ and one of its generators is $g^{\frac{n}{d}}$.
\end{Lemma}

\begin{Definition}[Coset]
Let us consider a subgroup $H\leq G$ of a commutative group $G$. A coset of $H$ is the set
\[gH=Hg=\{ gh : h \in H \}, \]
where $g\in G$.
\end{Definition}

It is straightforward to verify that $|gH|=|H|$ and that $gH=H$ if and only if $g\in H$.

\begin{Theorem}\label{cycgr}
Let $\mathbb{F}$ be a field. Then the multiplicative group \[\mathbb{F}^*=\mathbb{F}\backslash \{ 0 \}\]
is a cyclic group.
\end{Theorem}

\subsection{Elliptic Curves and the \textit{secp256k1}}
\label{curvesection}

\begin{Definition}[Elliptic Curve]
Let $\mathbb{F}$ be a field with characteristic different from $2$ and $3$. Let $A,B\in\mathbb{F}$ such that $\Delta=4A^3+27B^2\neq 0$. Then we define an elliptic curve $E(\mathbb{F})$ as the following subset of the affine plane over $\mathbb{F}$:
\[E(\mathbb{F})=\{(x,y)\in\mathbb{F}\times\mathbb{F} : \  y^2=x^3+Ax+B\}\cup \{\mathcal{O}\},\]
where $\mathcal{O}$ denotes the point at infinity.
\end{Definition}

It is possible to give an additive group structure to the set of points of an elliptic curve $E(\mathbb{F})$, with the inner chord-tangent point-addition (see Chapter 2 of \cite{WASH} for more details).\\

Let us now introduce the most relevant elliptic curve for our purposes, which is the one used by all the cryptocurrencies that we are going to consider.

\begin{Definition}[The secp256k1 curve]
Let us consider the prime number
\[p=2^{256}-2^{32}-2^9-2^8-2^7-2^6-2^4-1.\]
The elliptic curve
\[E(\mathbb{F}_p): \ Y^2=X^3+7 \]
is called \textit{secp256k1}.
\end{Definition}
\vskip0.5cm
The number of rational points of the curve \textit{secp256k1} over $\mathbb{F}_p$ is the prime number
\begin{align*}
q=&1157920892373161954235709850086879078528375642790749043\\
&82605163141518161494337,
\end{align*}
so that the additive group $\mathcal{E}=E(\mathbb{F}_p)$ is isomorphic to $\mathbb{F}_q (+)$. A generator of this group is the point $P=(P_x,P_y)$, where
\begin{align*}
P_x=&55066263022277343669578718895168534326250603453777594\\
&175500187360389116729240,\\
P_y=&32670510020758816978083085130507043184471273380659243\\
&275938904335757337482424.
\end{align*}

If they are given the point $P$ and a point $Q=kP$ (obtained by summing $k$ times the point $P$), it is in general very hard to recover the integer $k$. This is known as the Discrete Logarithm Problem over Elliptic Curves (ECDLP). 

\vskip1cm
\section{The small subsets}
As we have seen in Section \ref{sect2}, the group $E(\mathbb{F}_p)$ of the curve \textit{secp256k1} has prime order
\begin{align*}
q=&1157920892373161954235709850086879078528375642790749043\\
&82605163141518161494337,
\end{align*}
and it is then isomorphic to the additive group $\mathbb{F}_q(+)$. Hence, the private key can be chosen among the non-zero elements of $\mathbb{F}_q (+)$, which are $q-1$. The set of all private keys, then, is in bijection with the multiplicative group $\mathbb{F}_q^* (\cdot)$, which also has order $q-1$ and, by Theorem \ref{cycgr}, is a cyclic group. Therefore, by Lemma \ref{cyclic}, it has a unique cyclic subgroup of order $d$ for any divisor $d$ of $q-1$. Its factorisation is
\[q-1=h\cdot p_1 \cdot p_2 \cdot p_3,\]
where
\begin{align*}
h&=18051648=2^6 \cdot 3 \cdot 149 \cdot 631,\\
p_1&=107361793816595537,\\
p_2&=174723607534414371449,\\
p_3&=341948486974166000522343609283189.
\end{align*}
In \cite{SALA}, the authors examined the multiplicative subgroup $H\leq \mathbb{F}_q^*$, whose order is 
\[|H|=h=18051648,\]
since it can be fully investigated in a short time.
As pointed out in the same paper, it is worth to analyse also some other structures of the same order, for example the cosets of $G$.
Let us consider the generator $g=7$ of the group $\mathbb{F}_q^* (\cdot)$. In this way, using Lemma \ref{cyclic} and denoting by 
\begin{align*}
g_0&=7^{p_1p_2p_3}, \ & g_1&=7^{hp_2p_3}, \ &  g_2&=7^{hp_1p_3}, \ & g_3&=7^{hp_1p_2},\\
g_4&=7^{hp_1}, & g_5&=7^{hp_2}, & g_6&=7^{hp_3}, & g_7&=7^h,
\end{align*}
we have that
\begin{align*}
|\langle g_0\rangle |&=|H|=h, \ &  |\langle g_1\rangle |&=p_1, \ &  |\langle g_2\rangle |&=p_2,\\
|\langle g_3\rangle |&=p_3, \ & |\langle g_4\rangle |&=p_2p_3, \ & |\langle g_5\rangle |&=p_1p_3,\\ 
|\langle g_4\rangle |&=p_1p_2, \ & |\langle g_7\rangle |&=p_1p_2p_3.
\end{align*}
The cosets that we investigate in this paper are $g_iH$, for $i\in \{0,\ldots,7 \}$. Notice that $g_0H=H$, that is the same subgroup considered in \cite{SALA}.

\begin{Remark}
What the authors have done in \cite{SALA} is to search among the multiplicative subgroups of $\mathbb{F}_q^*(\cdot)$, even though the private key space is $(E(\mathbb{F}_p),+)\cong \mathbb{F}_q(+)$, hence additive. This observation makes the result ($4$ keys detected) even more surprising, since it has been considered a subgroup that is far from the actual nature of the group $E(\mathbb{F}_p)$. This very curious fact makes us think that the cause is some implementation error that has occurred during the practical design of the wallets.
\end{Remark}

\vskip1cm
\section{Address generation}
In this section we examine how the addresses are generated starting from the choice of the private key.
\subsection{Private and Public Key}
The private key is an arbitrary $256$-bit integer $k$. Using $k$, it is possible to compute the point
\[K=kP=(K_x,K_y)\]
over the elliptic curve, from which the public key is derived.
We have two different ways to represent the public keys:
\vskip0.3cm
\begin{enumerate}
    \item[i)] $PK_1=0\text{x}04 \,|| \,K_x \,|| \,K_y$,
    \vskip0.3cm
    \item[ii)] $PK_2=
    \begin{cases}
    0\text{x}02 \,|| \,K_x \ \  \text{if} \ K_y \ \text{is even},\\
    0\text{x}03\, || \, K_x \ \ \text{if} \ K_y \ \text{is odd},
    \end{cases}$
\end{enumerate}
\vskip0.2cm
where we have denoted with $||$ the string concatenation. In the first case, the public key, also known as \textit{uncompressed public key}, consists of $65$ bytes ($32$ bytes for $K_x$ and $K_y$, plus the byte $0\text{x}04$), while in the second case the public key is called \textit{compressed public key} and consists of $33$ bytes. The ways of obtaining the addresses for each cryptocurrency are discussed in the following paragraphs.

\vskip0.75cm
\subsection{Bitcoin addresses}
There are three manners to generate Bitcoin addresses starting from the point $K=(K_x,K_y)$.\\
First of all, two hash functions appear in Bitcoin address generation, which are SHA-256 \cite{SHA} and RIPEMD-160  \cite{RIPEMD}. Let us compute, for $i\in \{1,2\} $ :
\begin{align*}
W_i=0\text{x}00 \,||\,\text{RIPEMD-160}(\text{SHA-256}(PK_i)),\\
checksum_i=(\text{SHA-256}(\text{SHA-256}(W_i)))[1..4],
\end{align*}
where $[1..4]$ denotes the first $4$ bytes of that string. The first byte $0\text{x}00$ is a prefix called \textit{version byte}. 
Then, the first two ways to generate  Bitcoin addresses are, for $i\in \{1,2\} $,
\[\text{Base58}(W_i\, || \, checksum_i),\]
where Base58 \cite{Base58} is an encoding scheme.\\
The addresses of the third kind have been introduced in 2017 and they are called \textit{segwit} addresses \cite{segwit}. These addresses are computed only starting from compressed public keys. First of all, it is computed
\[W=\,\text{RIPEMD-160}(\text{SHA-256}(PK_2)),\]

then it is encoded using Bech32 \cite{bech32} and it is concatenated to the prefix \textit{bc1} in order to obtain the final address format
\[ bc1 \, || \, \text{Bech32}(W). \]

\subsection{Ethereum addresses}
Ethereum addresses are generated starting from the uncompressed public key, that is
\[PK=0\text{x}04||K_x||K_y.\]
However, a different hash function is used: KECCAK-256 \cite{KECCAK}. The address is computed as
\[\text{KECCAK-256}(PK)[1..20],\]
where $[1..20]$ denotes the first $20$ bytes of that string.\\
After this computation, the addresses are encoded following the rules described in the EIP-55 document \cite{eip55}. In short, the capitalisation of certain alphabetic characters in the address is changed to obtain a checksum that can be used to protect the integrity of the address from typing or reading errors.

\subsection{Dogecoin addresses}
Dogecoin address generation is similar to the first two methods described for Bitcoin, thus we can use uncompressed or compressed public keys. It only changes the \textit{version byte} $0\text{x}00$ of Bitcoin into $0\text{x}1\text{E}$. In this way, the final Base58 encoding gives a D as the first letter for each address. Dogecoin does not generate addresses following the \textit{segwit} standard.

\subsection{Litecoin addresses}
 Litecoin can generate addresses using all the three methods described for Bitcoin. In the first two cases, the prefix is the \textit{version byte} $0\text{x}30$ instead of $0\text{x}00$. In this way, all the addresses start with the letter $L$.
In the segwit case, the Bech32 encoding of the address is concatenated with the prefix $ltc1$, that is 
\[ltc1 \, || \, \text{Bech32}(\text{RIPEMD-160}(\text{SHA-256}(PK_2))).\]

\subsection{Dash addresses}
The address generation is similar to Bitcoin, but in this case the \textit{version byte} is changed into 0x4c, so that all the addresses start with the letter $X$. Dash does not generate addresses following the segwit standard.

\subsection{Zcash addresses}
The address generation is similar to Bitcoin, but in this case the \textit{version byte} is changed into [0x1c, 0xb8]. Zcash addresses can either start with the letter $t$, if they are transparent, or $z$, if they are shielded. Zcash does not generate addresses following the segwit standard.

\subsection{Bitcoin Cash addresses}
Bitcoin Cash (BCH) is the result of a  Bitcoin hard fork that happened in 2017, when some of Bitcoin nodes did not share the segwit soft fork. BCH addresses are encoded two times:
\begin{itemize}
    \item first, an address with the same encoding of Bitcoin is obtained,
    \item finally, the address is encoded again, with an encoding scheme called \textit{CashAddr} \cite{bchadd}, which is used by Bitcoin Cash only. This encoding is similar to Bech32.
\end{itemize}
BCH addresses that are encoded twice start with the string \textit{bitcoincash:q}, or just with the letter \textit{q}.
It comes natural that Bitcoin Cash does not generate addresses following the segwit standard.

\section{Experimental environment and development}
In this section we describe our experimental environment, which led us to the results showed in the following section. In order to perform a deep investigation for the existence of the addresses generated, we used a database filled with all addresses that had ever appeared on the blockchains of our interest, exploiting the MySQL \textit{Laragon} tool for the search operation  \cite{lar}.
\vskip0.5cm
\subsection{Blockchain addresses extraction}
The extraction of all the addresses ever appeared  on a blockchain can be done in different ways, depending on the time availability.\\
Generally, the two most common methods are: 
\begin{itemize}
  \item setting up a \textit{full node} (which contains a complete local copy of the relative blockchain) and reading data from it,
  \item getting data through public Application Programming Interfaces (APIs) (available on the web).
  \end{itemize}
In our case, we decided to set up full nodes when it was impossible to retrieve data from any public API. In this regard, we would like to point out that the availability of public APIs, documentation and general support were very weak for any blockchain other than Bitcoin and Ethereum.
\vskip0.5cm

\subsection{Development}
For some blockchains (Dogecoin, Litecoin, Bitcoin Cash and Ethereum) we developed some Python scripts in order to use the public APIs provided by \textit{Tatum} \cite{tat} and \textit{Ankr} \cite{ankr}. Differently, to analyse Dash and Zcash, we had to build our own full nodes and to get the blocks data from them. Finally, the list of Bitcoin addresses was taken from \cite{BS}, which offers some interesting information about the Bitcoin blockchain.
For anyone interested in learning more, our code is publicly accessible at \cite{code}. We conclude this section by reporting in the following table the number of addresses we have examinated, that is the totality of addresses that have ever been used in the history of these blockchains.
\vskip0.4cm

\begin{table}[!ht]
\centering
\begin{tabular}{|l|l|l|}
\hline
    Blockchain    &  Addresses \\
\hline 
  Bitcoin & 923.414.052 \\
   \hline 
Ethereum & 144.596.346  \\
\hline
 Dogecoin  &  69.817.509 \\
\hline
 Litecoin  &    134.530.241      \\
\hline 
 Dash  &  92.456.113        \\
\hline 
 Zcash  &    6.813.058     \\
\hline 
 Bitcoin Cash  &    334.965.092    \\
\hline 
\end{tabular}

\end{table}

\section{Cosets examination}
In this section, we list the results that we have obtained by inspecting the eight cosets $g_iH$, $i\in \{0,\ldots,7 \}$, of order $h=18051648$ for the seven blockchains we have chosen to investigate. The eight cosets contain a total of $8h \approx 144$ million addresses. We have checked if these addresses ever appeared in the blockchains mentioned above, i.e. if they have ever held any amount of cryptocurrency. 
\vskip0.25cm
More specifically, we have written some Python code that, starting from the list of private keys, computes the resulting addresses in the right format, and then checks if these addresses ever appeared on the analysed blockchains. The code used in this analysis is publicly accessible at \cite{code}. The following table sums up all the results.
\vskip0.5cm
\begin{table}[!ht]
\centering
\begin{tabular}{|l|l|l|}
\hline
        &  $H$& $g_iH$, $i=1,\ldots,7$\\
\hline 
   Uncompressed Bitcoin addresses & $4$ addresses found  \cite{SALA} & No address found \\
   \hline 
   Compressed Bitcoin addresses & $3$ addresses found & No address found \\
   \hline 
   Segwit Bitcoin addresses & $1$ address found & No address found \\
   \hline
Ethereum & 1 address found &  No address found \\
\hline
 Dogecoin  & 3 addresses found & No address found   \\
\hline
 Litecoin  & 2 addresses found & No address found         \\
\hline 
 Dash  & 2 addresses found & No address found         \\
\hline 
 Zcash  & No address found & No address found        \\
\hline 
 Bitcoin Cash  & 6 addresses found & No address found        \\
\hline 
\end{tabular}

\end{table}

From this analysis, it turns out that only three addresses are non-trivial, i.e. they do not have 1 or -1 as the private key. Notice that, for example, in Dogecoin we have found $3$ trivial addresses since each private key can be associated to two different encodings of the public key (compressed and uncompressed), for a total of $4$ potential trivial addresses. All of the three non-trivial addresses belong to the Bitcoin blockchain. The two addresses already found by Sala et al. \cite{SALA} were generated in 2013 and 2014, so they are present in Bitcoin Cash as well, while the third one is the address
\begin{center}
 1H1jFxaHFUNT9TrLzeJVhXPyiSLq6UecUy,
\end{center} 
and it was generated starting from a compressed public key. It was created on October 15, 2019, after the Bitcoin Cash fork, which is dated August 1, 2017. The address has no cryptocurrency into it nowadays.
\vskip0.25cm
The story of the address
\begin{center}
1PSRcasBNEwPC2TWUB68wvQZHwXy4yqPQ3,    
\end{center}
already found by Sala et al., is somewhat interesting. That address was created on March 15, 2014, and the funds got removed in June 2018 only after the authors of the previously cited papers contacted them (read \cite{SALA} to know more about what they have done). Hence, the address is old enough to appear on BCH. It is indeed present:
\begin{center}
qrmzrdndlfxpnkk3w5d5l7etnysnqfgk5yxsf6k0qq,
\end{center}
after the new encoding with the CashAddress format.\\
However, the address is empty as someone moved the funds on that address on May 1, 2019. Then on November 15, 2018, Bitcoin Cash had yet another hard fork, that led to the birth of \textit{Bitcoin SV}. Hence, this address must be present on that blockchain as well, with the funds still into it. By examining a Bitcoin SV explorer, it turns out that the funds from that address got also moved on May 1, 2019, two minutes earlier than the transaction on Bitcoin Cash! It seems likely to assume that the funds got moved from the same entity. 
\vskip0.5cm
Finally, notice that we were not able to find any address on the seven cosets we have chosen to examine. This is interesting, but not unexpected, as the probability to find an address with this brute-force method is really low \cite{SALA}. In this way, the security of these blockchain is not threatened since it results way more profitable to mine, i.e. to join honestly the protocol, than trying to generate random private keys with the aim of stealing cryptocurrency \cite{MINE}.
This may also confirm that the non-trivial addresses were generated due to a poor implementation of some wallets. It would also be interesting to try to understand which Bitcoin wallets generated these addresses.

\subsection{Other curves to examine}\label{seccur}
The most used elliptic curve in\\ blockchains, beside the \textit{secp256k1}, is for sure the \textit{Curve25519}. It is employed in several blockchains, including Monero \cite{Mon}, Cardano \cite{Car}, Solana \cite{Sol} and Algorand \cite{Alg}.
The defining curve is the Montgomery curve
\[E(\mathbb{F}_p): \ Y^2=X^3+486662X^2+X,\]
where $p=2^{255}-19$ is a prime number. The number of rational points is $n=8l$ where
\[l=2^{252} + 27742317777372353535851937790883648493.\]
Reasoning as in the case of the curve \textit{secp256k1}, the private key can be chosen among the non-zero points, whose number is
\[l-1=2^2\cdot 3 \cdot 11 \cdot q_1 \cdot q_2,\]
where
\begin{align*}
q_1&=276602624281642239937218680557139826668747,\\
q_2&=198211423230930754013084525763697.
\end{align*}
The private key space is then in bijection with $\mathbb{Z}_{l}^*(\cdot)$, it is cyclic and has a unique subgroup for each divisor of its order $l-1$, by Lemma \ref{cyclic}. Notice that an analogue analysis can not be performed, since the small subgroup contains only $132=2^2\cdot 3 \cdot 11$ points.\\

Finally, another curve that might be worth analysing in the future is the NIST \textit{P-256} curve, defined in the specification \cite{NIST}. It  is actually used by NEO \cite{NEO}, Tezos \cite{TEZ} and Ontology \cite{ONT}. The curve is
\[E(\mathbb{F}_p): \ Y^2 = X^3-3X+B,\]
where
\begin{align*}
B=&410583637251521421293261297800472684091144410159937255\\
&54835256314039467401291,
\end{align*}
and $p$ is the prime number
\[p=2^{256}-2^{224}+2^{192}+2^{96}-1.\]
The order of this curve is
\begin{align*}
q=&115792089210356248762697446949407573529996955224135760\\
&342422259061068512044369,
\end{align*}
and $q-1$ factorises as
\[q-1=2^4\cdot 3\cdot 71 \cdot 131 \cdot 373\cdot 3407 \cdot 17449\cdot 38189\cdot 187019741\cdot 622491383\cdot q_1\cdot q_2,\]
where
\begin{align*}
q_1&=2624747550333869278416773953,\\
q_2&=1002328039319.
\end{align*}
This case is certainly more interesting than the previous one, since there are several subgroups of order dividing $q-1$ that can be inspected.

\section{Conclusions}
In this paper we have tried to give an answer to two open questions left in \cite{SALA}. It turns out that the strange behaviour observed by Sala et al. is  a peculiarity of Bitcoin (and, of course, of its hard forks), where we have also found a new address not detected in \cite{SALA}, since it uses the compressed public key format. Instead, we have not been able to find any non-trivial private keys for the other analysed blockchains and for the cosets of the starting subgroup. \\ \ \\
Another contribution of this work is the comprehensive analysis of the addresses we have extracted from some of the most important blockchains to date.  In fact, it is almost impossible to find relevant information on the web for any blockchain other than Bitcoin and, for some extent, Ethereum. The interested readers can extract addresses with the same methods, using the scripts available in our GitHub folder \cite{code}.\\ \ \\
For future works, it might be interesting to analyse more thoroughly the wallets used in the period these addresses were generated, to try to confirm the hypothesis that this behavior really comes from an incorrect implementation of the wallet itself. In addition, it would be possible to examine the small subgroups found in Section \ref{seccur} to understand if also NIST $P-256$ curve presents some strange behaviour on those subsets.

\end{document}